\documentclass[usenatbib,fleqn,useAMS,twocolumn]{mnras}
\usepackage[dvipdfmx]{graphicx}
\usepackage[dvipdfmx]{color}
\usepackage{amsmath,amssymb}
\usepackage{epstopdf}
\usepackage{grffile}
\usepackage{times}
\usepackage{url}
\usepackage{bm}
\usepackage{xcolor}

\newcommand{\simgt}{\lower.5ex\hbox{$\; \buildrel > \over \sim \;$}}
\newcommand{\simlt}{\lower.5ex\hbox{$\; \buildrel < \over \sim \;$}}

\newcommand{\himpc}{{\hbox {$~h^{-1}$}{\rm ~Mpc}}}

\newcommand{\vx}{\mathbf{x}}

\newcommand{\vk}{\mathbf{k}}

\newcommand{\vv}{\mathbf{v}}
\newcommand{\vecr}{\mathbf{r}}
\newcommand{\hvk}{\hat{\mathbf{k}}}

\newcommand{\be}{\begin{equation}}
\newcommand{\ee}{\end{equation}}
\newcommand{\bey}{\begin{eqnarray}}
\newcommand{\eey}{\end{eqnarray}}

\newcommand{\wt}{\widetilde}

\newcommand{\nn}{\nonumber}

\newcommand{\paper}{Letter} 

\begin{document}
\title[Anisotropies of galaxy ellipticity correlations]{Anisotropies of galaxy ellipticity correlations in real and redshift space: 
angular dependence in linear tidal alignment model
}

\author[T. Okumura and A. Taruya]{
\parbox{\textwidth}{
Teppei Okumura$^{1,2}$\thanks{tokumura@asiaa.sinica.edu.tw} and Atsushi Taruya$^{3,2}$}
\vspace*{4pt} \\
$^{1}$ Institute of Astronomy and Astrophysics, Academia Sinica,
No. 1, Section 4, Roosevelt Road, Taipei 10617, Taiwan \\
$^{2}$ Kavli Institute for the Physics and Mathematics of the Universe (WPI), 
UTIAS, The University of Tokyo, Kashiwa, Chiba 277-8583, Japan \\
$^{3}$ Center for Gravitational Physics, Yukawa Institute for Theoretical Physics, Kyoto University, Kyoto 606-8502, Japan}

\date{Accepted 2020 February 6. Received 2020 February 4; in original form 2020 December 16} 
\pagerange{\pageref{firstpage}--\pageref{lastpage}} \pubyear{2019}

\maketitle
\label{firstpage}

\begin{abstract}
Investigating intrinsic alignments (IAs) of galaxy shapes is important not only to constrain cosmological parameters unbiasedly from gravitational lensing but also to extract cosmological information complimentary to galaxy clustering analysis. We derive simple and useful formulas for the various IA statistics, including the intrinsic ellipticity--ellipticity correlation, the gravitational shear--intrinsic ellipticity correlation, and the velocity-intrinsic ellipticity correlation functions. The angular dependence of each statistic is explicitly given, namely the angle between the line-of-sight direction and the separation vector of two points. It thus allows us to analyze anisotropies of baryon acoustic oscillations encoded in the IA statistics, and we can extract the maximum cosmological information using the Alcock-Paczynski and redshift-space distortion effects. We also provide these formulas for the intrinsic ellipticities decomposed into $E$ and $B$ modes. \end{abstract}
\begin{keywords}
methods: statistical -- galaxies: haloes -- cosmological parameters -- cosmology: theory -- dark energy -- large-scale structure of universe.
\end{keywords}


\section{Introduction} \label{sec:intro}

Intrinsic alignments (IAs) of galaxy orientations/shapes are known as a contamination to cosmological parameter estimations with the weak lensing surveys \citep{Heavens:2000,Croft:2000,Catelan:2001,Crittenden:2001,Hirata:2004} \citep[See e.g.][for a review]{Troxel:2015}. 
In weak lensing surveys, two types of IAs have been considered: the ellipticity correlation of source galaxies with each other (intrinsic ellipticity--ellipticity (II) correlation) and the ellipticity correlation of lens galaxies with the surrounding matter distribution (gravitational shear--intrinsic ellipticity (GI) correlation). \cite{Hirata:2004} showed that the prediction of the linear alignment (LA) model \citep[][see Sec.~\ref{sec:la} below]{Catelan:2001} provides formulas for the II and GI correlations in Fourier space and they are simply proportional to the linear power spectrum in the large-scale limit. While the LA model provides excellent agreement with the various alignment correlations in simulations and observations even in configuration space, these statistics were expressed in a more complex form than in Fourier space, including a double integral over $k_\perp$ and $k_\parallel$, which are the wavenumbers perpendicular and parallel to the line of sight ($k^2=k_\perp^2+k_\parallel^2$) \citep{Blazek:2011,Okumura:2019}. Thus, the angular dependence of the statistics is not as trivial as the Fourier-space counterpart,

The purpose of this \paper\, is to perform one of the double integral analytically and provide the intrinsic alignment statistics in configuration space with simpler forms. It means that angular dependences of all the 2-point alignment statistics in three-dimensional space can be explicitly given by the form of $\xi(\vecr)=\xi(r,\mu)$, where $\mu$ is the direction cosine between the line of sight and separation vector $\vecr$. By construction, anisotropies arise in the alignment correlation statistics even in real space because observable shapes of galaxies are the projection along the observer's line of sight. However, in the literature only the angle-averaged quantities (monopole) have been considered, and thus the IA signal has not fully been explored. Revealing the angular dependence allows one to extend the statistics straightforwardly from real space to redshift space where galaxies are actually observed \citep{Kaiser:1987, Hamilton:1998}. Recently there are several studies which utilize IAs as a complimentary cosmological probes \citep{Schmidt:2012,Chisari:2013,Chisari:2014,Schmidt:2015,Chisari:2016,Kogai:2018}. The derived formulas in this work will be essential to extract the full cosmological information from IAs by utilizing the Alcock-Paczynski effect \citep{Alcock:1979,Taruya:2020}. We will also derive a formula for the GI correlation in phase space, namely the velocity field-intrinsic ellipticity (VI) correlation \citep{Okumura:2019} which can be measured by the kinematic Sunyaev-Zel'dovich surveys \citep{Sunyaev:1980}. 

\section{Intrinsic alignment statistics} \label{sec:theory}
In this section we briefly describe the statistics used to characterize IAs. 

First, the two components of the ellipticity of each galaxy (or cluster) are given as 
\be
\gamma_{(+,\times)}(\vx)=\frac{1-(\beta/\alpha)^2}{1+(\beta/\alpha)^2}(\cos(2\theta),\sin(2\theta)), \label{eq:gamma_pc}
\ee
where $\beta/\alpha$ is the minor-to-major axis ratio, $\theta$ is the position angle of the ellipticity defined on the plane normal to the line-of-sight direction, and the ellipticity is also defined on the projected plane (see fig.~1 of \cite{Okumura:2019} for the illustration of these quantities, and note $\theta\neq \cos^{-1}{\mu}$). Sometimes the superscript $I$ is added to $\gamma_{+,\times}$ to distinguish intrinsic ellipticities from the cosmic shear components in weak lensing surveys. However, we omit it because lensing is not considered in this \paper.

The II correlation of galaxies has four components, and one of the four, $\xi_{++}$, is defined as \citep{Heavens:2000, Croft:2000}
\begin{align}
1+\xi_{++}(\vecr)=\left\langle[1+\delta_{\rm g}(\vx_1)][1+\delta_{\rm g}(\vx_2)]\gamma_+(\vx_1)\gamma_+(\vx_2)\right\rangle, \label{eq:ii}
\end{align}
where $\vecr=\vx_2-\vx_1$. The other components, such as $\xi_{\times\times}$ and $\xi_{+\times}$, are defined in the same way by replacing two and one $\gamma_+$ in equation (\ref{eq:ii}) with $\gamma_\times$, respectively. By combining $\xi_{++}$ an $\xi_{\times\times}$, we can also define $\xi_\pm(\vecr)$ as
\be 
\xi_\pm(\vecr) = \xi_{++}(\vecr) \pm \xi_{\times\times}(\vecr). 
\ee

The cross-correlation functions of density and ellipticity fields, namely GI correlations, are defined as \citep{Hirata:2004}
\be
1+\xi_{{\rm g}i}(\vecr)=\left\langle[1+\delta_{\rm g}(\vx_1)][1+\delta_{\rm g}(\vx_2)]\gamma_i(\vx_2)\right\rangle, \label{eq:gi}
\ee
where $i=\{+,\times \}$. Since the distances to objects are measured through redshift in galaxy surveys, the density field is affected by their velocities, known as redshift-space distortions (RSDs) \citep{Kaiser:1987,Hamilton:1998}. Thus, the superscripts $R$ and $S$ are added to $\xi_{{\rm g}+}$ to denote the GI correlation in real and redshift space, respectively.

We also consider the velocity alignment statistic corresponding to the GI correlation, the density-weighted, velocity-intrinsic ellipticity (VI) correlation \citep{Okumura:2019},
\begin{align}
\xi_{vi}(\vecr)=\left\langle[1+\delta_{\rm g}(\vx_1)][1+\delta_{\rm g}(\vx_2)] v_\parallel(\vx_1)\gamma_i (\vx_2) \right \rangle, \label{eq:vi}
\end{align}
where $i=\{+,\times\}$ and $v_\parallel$ denotes the line-of-sight component of the velocity field, $v_\parallel(\vx)\equiv \vv(\vx)\cdot \hat{\vx}$ (hat denotes a unit vector). As is the case with the ellipticity field, the velocity field is not affected by RSDs in linear theory, $\xi_{v+}^S=\xi_{v+}^R$ \citep{Okumura:2014,Okumura:2017a}. 

All the statistics above are anisotropic even in real space because observable shapes of galaxies are the line-of-sight projection. Moreover, RSDs induce further anisotropies to the the GI correlation function. Thus, we consider the multipole moments of the correlation functions \citep{Hamilton:1992}:
\be
X_{{\ell}}(r)=\frac{2\ell+1}{2}\int^1_{-1} d\mu X(\vecr){\cal P}_\ell(\mu), 
\ee
where $X$ is any of the statistics introduced above, and
$\mu$ is the directional cosine between the vector $\vecr$ and the line-of-sight direction $\hat{\vx}$. Below, we use $r_\perp$ and $r_\parallel$ to express respectively the separations perpendicular and parallel to the line-of-sight direction. These are related to $r$ and $\mu$ through $r^2=r_\perp^2+r_\parallel^2$ and $\mu=r_\parallel/r$. Throughout this \paper\, we assume the distant-observer approximation, and particularly take $z$-axis to be the line-of-sight direction so that $\hat{\vx}_1=\hat{\vx}_2\equiv\hat{\vx}$. 

\section{Linear alignment model}\label{sec:la}
The most commonly used model for IA studies on large scales is the LA model \citep{Catelan:2001,Hirata:2004}. In this model, the intrinsic ellipticity (equation \ref{eq:gamma_pc}) is assumed to follow the linear relation with the Newtonian potential, $\Psi_P$,
\be
\gamma_{(+,\times)}(\vx)= -\frac{C_1}{4\pi G}\left( \nabla_x^2-\nabla_y^2, 2 \nabla_x\nabla_y \right) \Psi_P(\vx), 
\label{eq:gamma_la}
\ee
where $G$ is the Newtonian gravitational constant, $C_1$ parameterizes the strength of IA. The observed ellipticity field is density-weighted, $[1+\delta_{\rm g}(\vx)]\gamma_{(+,\times)}(\vx)$ (Section \ref{sec:theory}). However, the density-weighting term $\delta_{\rm g}(\vx)\gamma(\vx)$ is sub-dominant on large scales and is usually ignored. We also do not consider this term because we are interested in the large-scale behaviors. In Fourier space, equation (\ref{eq:gamma_la}) becomes
\begin{align}
\gamma_{(+,\times)} (\vk) =  -\wt{C}_1 
\frac{\left(  {k}_{x}^{2}- {k}_{y}^{2}, 2{k}_x{k}_y \right)}{k^2}\delta(\vk),
\end{align}
where $\wt{C}_1(z)\equiv a^2 C_1\bar{\rho}(z)/\bar{D}(z)$, $\bar{\rho}$ is the mean mass density of the Universe, $\bar{D}\propto (1+z)D(z) $, and $D(z)$ is the linear growth factor.

The three-dimensional cross-correlation function between the density field and the ellipticity is given in the LA model as \citep{Okumura:2019} 
\begin{align}
\xi_{{\rm g}+ }(\vecr) 
= \ &  \wt{C}_1b_{\rm g} \cos{(2\phi)}\int^{\infty}_{0} \frac{k_\perp dk_\perp}{2\pi^2} J_2(k_\perp r_\perp) 
\nn \\
& \times \int^{\infty}_0 dk_{\parallel} \frac{k_\perp^2}{k^2} P_{\delta\delta}(k) \cos{(k_\parallel r_\parallel)} ,
\label{eq:gi_la}
\end{align}
where $k_\perp^2=k_x^2+k_y^2$, $k_\parallel = k_z$, $\phi$ is the azimuthal angle of the projected separation vector on the celestial sphere, measured from the $x$-axis, $J_2$ is the Bessel function with second order, $P_{\delta\delta}$ is the auto power spectrum of density and $b_{\rm g}$ is the linear galaxy bias parameter. Likewise, the II and VI correlation functions are expressed using the Bessel function (see \cite{Blazek:2011} and \cite{Okumura:2019}, respectively). Here and in what follows, we keep the $\phi$-dependence explicitly for clarity and completeness when a statistic is newly derived, and
we set $\phi=0$ when the multipole moments are further derived. 

\section{New formulas for IA statistics with linear alignment model}\label{sec:formula}
In this section we present formulas of the IA statistics, namely the GI, II and VI correlation functions in the LA model. We also show the results of the numerical calculations at $z=0.3$, for which we set the parameter $\wt{C_1}$ to $\wt{C_1}/a^2=1.5$, as determined by \cite{Okumura:2019} for dark matter haloes with the mass greater than $10^{14}\,M_\odot$. 

For later convenience, we newly introduce a quantity $\Xi_{XY,\ell}^{(n)}(r)$ defined by  
\be
\Xi_{XY,\ell}^{(n)}(r) = (aHf)^n\int^{\infty}_0 \frac{k^{2-n}dk}{2\pi^2}P_{XY}(k) j_\ell(kr), \label{eq:xi_formula}
\ee
where $XY=\{ \delta\delta, \delta\Theta, \Theta\Theta\}$, $\Theta$ is the velocity-divergence field defined by $\Theta(\vx)=-\nabla \cdot \vv/(aHf)$, $H(a)$ is the Hubble parameter and $f$ is the linear growth rate, given by $f\equiv d\ln D/d\ln a$. The quantities $P_{\delta\Theta}$ and $P_{\Theta\Theta}$ are the cross power spectrum of density and velocity divergence and the auto spectrum of the latter, respectively. In the linear theory limit, $P_{\delta\delta}=P_{\delta\Theta}=P_{\Theta\Theta}$.

\begin{figure*}
\includegraphics[width=0.3934\textwidth,angle=0,clip]{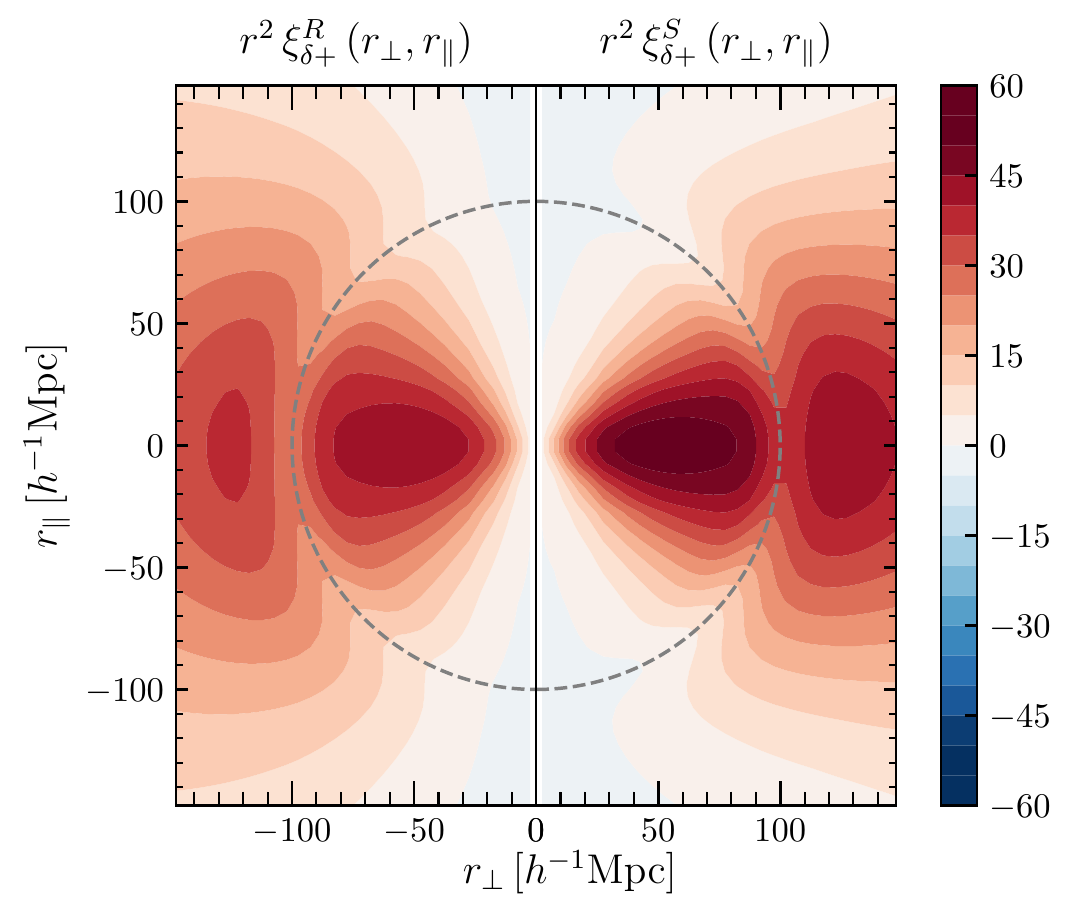}
\includegraphics[width=0.3724\textwidth,angle=0,clip]{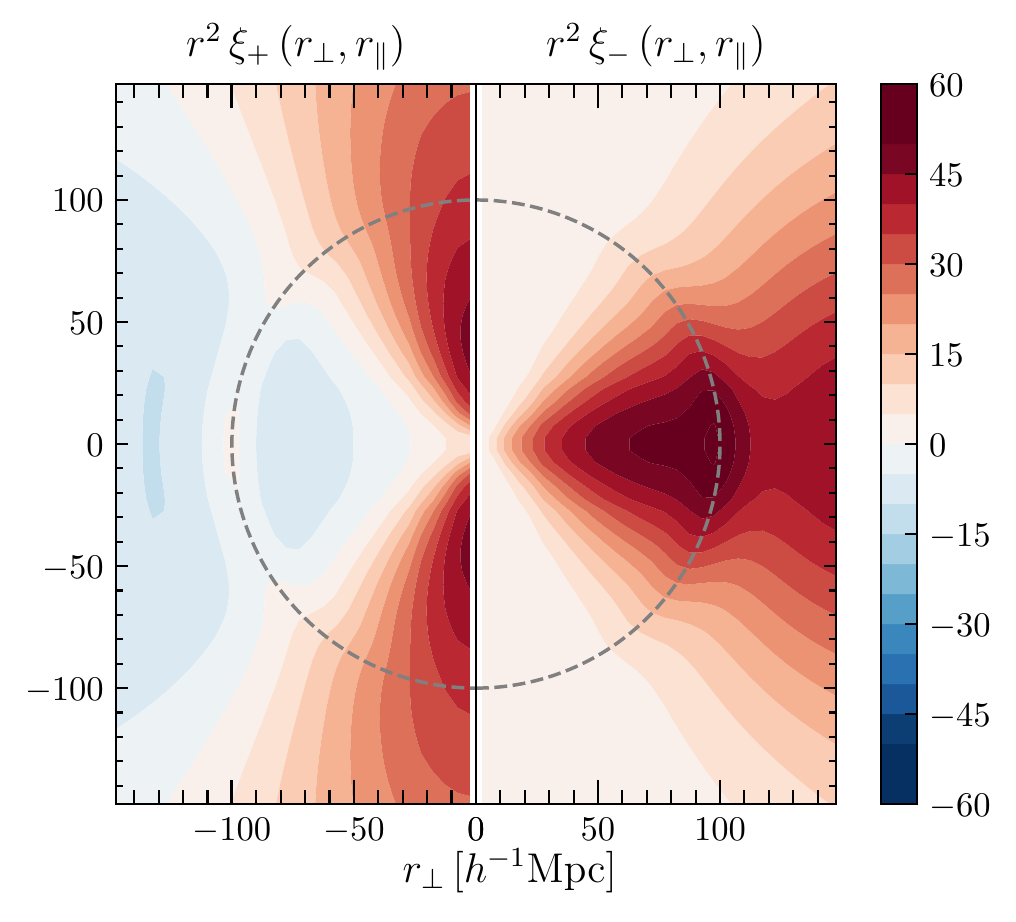}
\includegraphics[width=0.2109\textwidth,angle=0,clip]{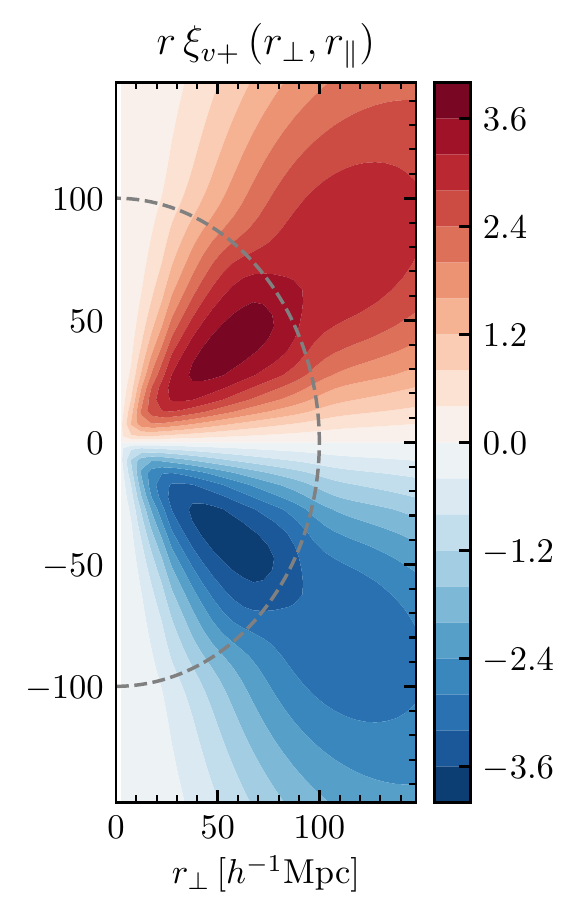}
\caption{
Left panel: GI correlation function as a function of separations perpendicular and parallel to the line of sight in real space $r^2 \xi_{\delta+}^R$ (left) and in redshift space $r^2 \xi_{\delta+}^S$ (right). The difference between the left and right hand sides is due to RSDs. Middle panel: Two II correlation functions, $r^2 \xi_{+}$ (left) and  $r^2 \xi_{-}$ (right). Right panel: VI correlation function $r \xi_{v+}$. The BAO scale, $r\simeq 100\himpc$, is denoted by the dashed gray circles in all the panels. All the statistics are calculated at $z=0.3$. 
}
\label{fig:contour2D_gi_ia_th_2gpc}
\end{figure*}

\subsection{GI correlation}

The conventional expression of alignment statistics in the LA model, such as equation (\ref{eq:gi_la}) for the GI correlation, was derived by adopting cylindrical coordinates. We rewrite all the angular dependences in Fourier space by the spherical harmonics, e.g. 
$(k_x^2-k_y^2)/k^2 = \sqrt{2/3}\, [y_{2,2}(\hvk)-y_{2,-2}(\hvk) ] $ 
where $y_{\ell m}(\hat{\vk}) \equiv \sqrt{4\pi/(2\ell +1)}Y_{\ell m}(\hat{\vk})$ is a normalized spherical harmonic function, and utilize its orthogonality condition. The angular integral then can be analytically performed. We find that the GI correlation function in real space is reduced to a much simpler form:
\begin{align}
\xi_{{\rm g}+ }^R(\vecr) 
=\wt{C}_1 b_{\rm g} 
\cos{(2\phi)}(1-\mu^2) 
\Xi_{\delta\delta,2}^{(0)}(r).
\label{eq:gi_la_1d}
\end{align}
This is equivalent to equation (\ref{eq:gi_la}), but here the angular dependence is explicitly given. Similarly, $\xi_{g\times}^R$ is described by replacing $\cos{(2\phi)}$ in equation (\ref{eq:gi_la_1d}) with $\sin{(2\phi)}$. 

The resulting GI correlation function as a function of $\vecr = (r_\perp,r_\parallel)$ is shown in the left half of the left panel in Fig. \ref{fig:contour2D_gi_ia_th_2gpc}. Here for simplicity we plot equation (\ref{eq:gi_la_1d}) with $b_{\rm g}=1$, which corresponds to the cross-correlation between matter density and galaxy ellipticity fields, $\xi_{\delta+ }^R(\vecr)=\xi_{{\rm g}+ }^R(\vecr)/b_{\rm g}$. The ridge structures seen around $r\simeq 100\himpc$ are the baryon acoustic oscillation (BAO) features \citep{Sunyaev:1970,Peebles:1970,Eisenstein:2005}. Similarly to the correlation function of the density field, the feature appears as a ``BAO ring'' \citep{Matsubara:2004,Okumura:2008}, but interestingly, it shows up as a dip in the GI correlation rather than a peak \citep{Okumura:2019}.

\begin{figure}
\includegraphics[width=0.47\textwidth,angle=0,clip]{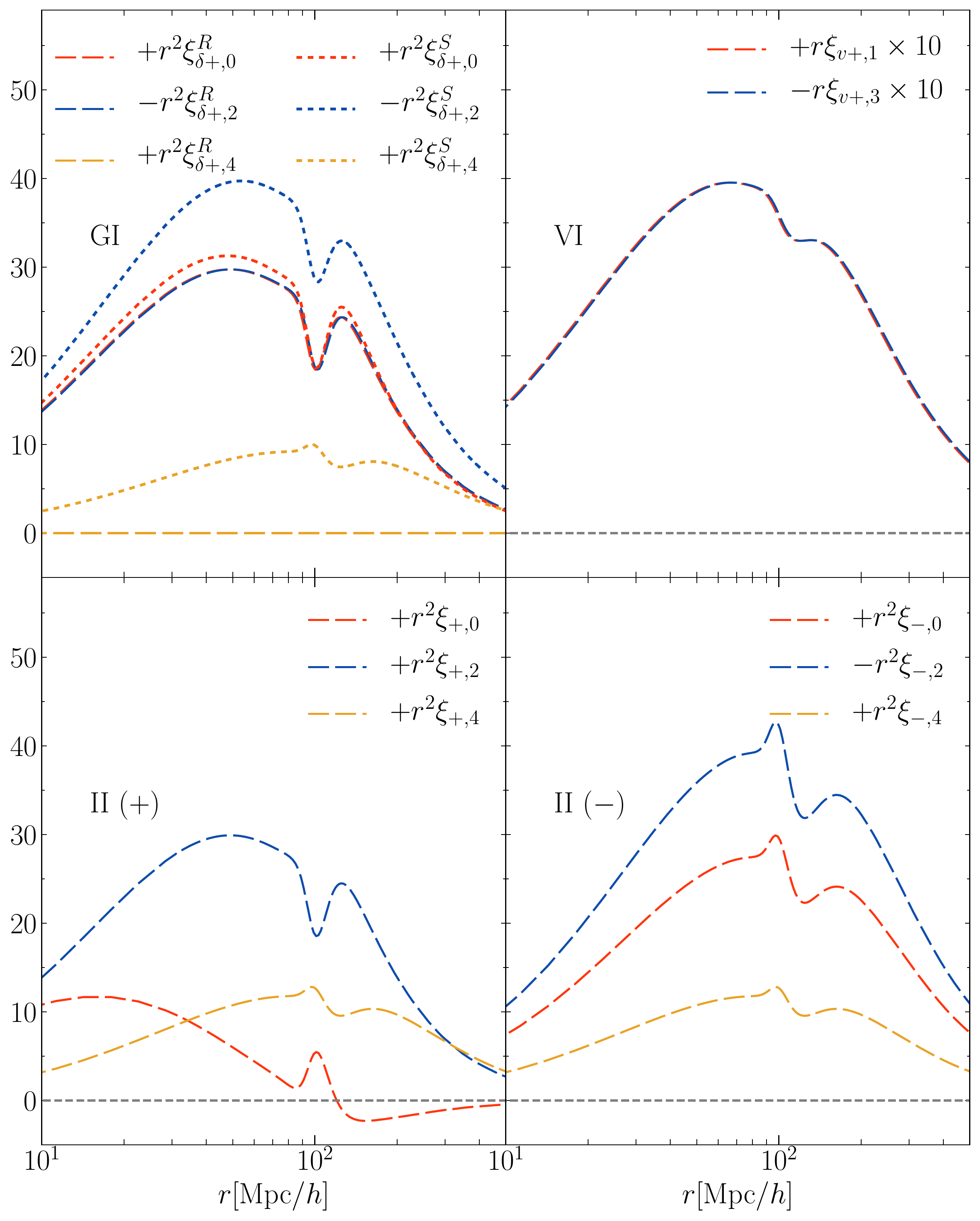}
\caption{
Multipole moments of correlation functions. The upper-left panel shows the GI correlation function in real space (dashed) and in redshift space (dotted), while the upper-right panel presents the VI correlation function. The bottom panels show the two components of the II correlation functions, $\xi_+$ (lower-left) and $\xi_-$ (lower-right). The GI and II correlations are multiplied by $r^2$, while the VI correlation is multiplied by $r$ and a factor of $10$. All the statistics are calculated at $z=0.3$. 
}
\label{fig:gi_ii_vi_multi_th_2gpc_z014}
\end{figure}

Obviously, the multipoles components of equation (\ref{eq:gi_la_1d}), $\xi_{{\rm g}+,\ell }^R(r)$, become non-zero only if $\ell = 0$ or $\ell=2$, and 
\be
\xi_{{\rm g}+,0 }^R(r) =-\xi_{{\rm g} +,2 }^R(r)  =\frac{2}{3}\wt{C}_1 b_{\rm g}\Xi_{\delta\delta,2}^{(0)}(r). \label{eq:gi_la_l}
\ee
This is shown as the red dashed curve in the upper-left panel of Fig.~\ref{fig:gi_ii_vi_multi_th_2gpc_z014}. It is equivalent with the red curve in fig.~2 of \cite{Okumura:2019}. The quadrupole-to-monopole ratio being $-1$ is a natural consequence of the LA model.

Next, let us extend the real-space formulation of the GI correlation to redshift space. We consider the Kaiser's RSD model \citep{Kaiser:1987}, $\delta_{\rm g}^S(\vk)=\delta_{\rm g}^R(k)+f\,(k_z/k)^2\Theta(k)$, where $\Theta(k)$ is the Fourier transform of the velocity divergence. We then have the additional angular-dependent term, $k_z^2/k^2 = \frac{2}{3}\,y_{2,0}(\hvk)+\frac{1}{3}\,y_{0,0}(\hvk)$. We can perform the integral using the relation between the spherical harmonics and Wigner's 3-j symbols, 
$\int d^2\hvk\, y_{\ell m}(\hvk)y_{\ell_1 m_1}(\hvk)y_{\ell_2 m_2}(\hvk)  =4\pi
\left( \begin{smallmatrix} \ell & \ell_1 & \ell_2 \\ 0 & 0 & 0 \\ \end{smallmatrix} \right)
\left( \begin{smallmatrix} \ell & \ell_1 & \ell_2 \\ m & m_1 & m_2 \\ \end{smallmatrix} \right)$.
The resulting GI correlation function in redshift space reads
\begin{align}
\xi_{{\rm g}+ }^S(\vecr) 
=&\xi_{\rm{g}+ }^R(\vecr) 
+\frac{1}{7}\wt{C}_1 f \cos{(2\phi)} \left(1-\mu^2\right) \nn \\
&\times\left[\Xi_{\delta\Theta,2}^{(0)}(r)-\left(7\mu^2-1\right)  \Xi_{\delta\Theta,4}^{(0)}(r)\right].
\label{eq:gi_la_s_1d}
\end{align}
The redshift-space GI correlation function is presented in the right half of the left panel of Fig.~\ref{fig:contour2D_gi_ia_th_2gpc}. Just like the density correlation function, RSDs do not shift the scale of BAO peak in the alignment correlation in linear theory. Thus, the alignment statistics can be used for the Alcock-Paczynski test complimentarily to the galaxy clustering statistics.

In redshift space, not only the monopole and quadrupole but also hexadecapole are the non-vanishing multipoles for the GI correlation function in the LA model:

\begin{align}
\xi_{{\rm g}+,0 }^S(r) 
&=\xi_{{\rm g}+,0 }^R(r) 
+\frac{2}{105}\, \wt{C}_1 f 
\left[5\,\Xi_{\delta\Theta,2}^{(0)}(r)- 2\,\Xi_{\delta\Theta,4}^{(0)}(r) \right], \\
\xi_{{\rm g}+,2 }^S(r) 
&=\xi_{{\rm g}+,2 }^R(r) 
-\frac{2}{21}\, \wt{C}_1\, f \, 
\left[\,\Xi_{\delta\Theta,2}^{(0)}(r)+ 2\,\Xi_{\delta\Theta,4}^{(0)}(r)\right], \\
\xi_{{\rm g}+,4 }^S(r) 
&=
\frac{8}{35}\, \wt{C}_1\, f \,
\Xi_{\delta\Theta,4}^{(0)}(r).
\end{align}
In the presence of the RSD effect, the quadrupole-to-monopole ratio is no longer $-1$ unlike the real-space case, and we have
$\xi_{{\rm g}+,2 }^S(r) / \xi_{{\rm g}+,0 }^S(r) < -1 $. 
These three multipole moments are shown as the dotted curves in the upper-left panel of Fig.~\ref{fig:gi_ii_vi_multi_th_2gpc_z014}. 

It is interesting to note that the quadrupole and hexadecapole moments of the redshift-space galaxy correlation function are given by \citep{Hamilton:1992} 
\begin{align}
\xi_{{\rm gg},2}^S(r)&=\frac{4}{3}\,f\,b_{\rm g} \Xi_{\delta\Theta,2}^{(0)}(r)+\frac{4}{7}\,f^2\,\Xi_{\Theta\Theta,2}^{(0)}(r), \\
\xi_{{\rm gg},4}^S(r)&=\frac{8}{35}\,f^2\, \Xi_{\delta\Theta,4}^{(0)}(r).
\end{align}
Namely, the GI correlation in real space has exactly the same shape as the quadrupole of the density correlation in redshift space in the linear theory limit, and likewise the GI correlation in redshift space can be described by the combination of the quadrupole and hexadecapole correlation functions. These features of the GI correlation function are clarified for the first time by our simple formulas.

\subsection{II correlation}
We can derive simple formulas for the II correlation in a similar way, although the II correlation function has a bit intricate form compared to  the GI correlation. The angular-dependent terms in $\xi_{++}$ and $\xi_{\times\times}$ are respectively rewritten as 
$\frac{1}{k^4}\left((k_x^2-k_y^2)^2, 4k_x^2k_y^2\right) = \pm \sqrt{\frac{8}{35}} \left [y_{4,4}(\hvk) + y_{4,-4}(\hvk)\right]
+\frac{4}{35}y_{4,0}(\hvk) - \frac{8}{21}y_{2,0}(\hvk) + \frac{4}{15}y_{0,0}(\hvk)$.
After applying the orthogonality condition of $y_{\ell m}$, the two components of the II correlation function, $\xi_\pm(\vecr)$, are  
given as \citep[see][for an similar expression for the monopole moment]{Xia:2017}
\begin{align}
\xi_{+}(\vecr) 
=&\frac{8}{105}\,\wt{C}_1^2
 \left[
7\,{\cal P}_0(\mu)\, \Xi_{\delta\delta,0}^{(0)}(r)
+10\,{\cal P}_2(\mu)\, \Xi_{\delta\delta,2}^{(0)}(r)
\right. \nn \\&  
\left. 
+3\,{\cal P}_4(\mu)\, \Xi_{\delta\delta,4}^{(0)}(r)
\right],
\label{eq:ii_la_1d_p} \\
\xi_{-}(\vecr) 
=&
~\wt{C}_1^2
\cos{(4\phi)} \left( 1-\mu^2\right)^2 \, \Xi_{\delta\delta,4}^{(0)}(r)
\nn 
 \\ 
 =&\frac{8}{105}\,
 \wt{C}_1^2\,
 \cos{(4\phi)} \nn \\
&  \times  \left[
7\,{\cal P}_0(\mu) +10\,{\cal P}_2(\mu) +3\,{\cal P}_4(\mu) 
\right]\,\Xi_{\delta\delta,4}^{(0)}(r).
\label{eq:ii_la_1d_m}
\end{align}
Since the II correlation function is not affected by RSDs in linear theory, $\xi_{\pm}^S=\xi_{\pm}^R$, we omit the superscript for this statistic. The cross component, $\xi_{+\times}$, can be obtained by replacing $\cos{(4\phi)}$ in equation (\ref{eq:ii_la_1d_m}) with $\sin{(4\phi)}$. The II correlations, $\xi_+$ and $\xi_-$, are respectively presented in the left and right hand sides of the middle panel of Fig.~\ref{fig:contour2D_gi_ia_th_2gpc}. Combining these two functions, one can also derive $\xi_{++}$ and $\xi_{\times\times}$, and our formula nicely explains the anisotropic feature of $\xi_{\times\times}$ measured from $N$-body simulations by \cite{Croft:2000}. 

The multipole components of $\xi_\pm(\vecr)$ are obvious from equations (\ref{eq:ii_la_1d_p}) and (\ref{eq:ii_la_1d_m}), and their hexadecapoles coincide with each other. The resulting multipoles, $\xi_{+,\ell}$ and $\xi_{-,\ell}$, are respectively shown in the lower-left and lower-right panels of Fig. \ref{fig:gi_ii_vi_multi_th_2gpc_z014}. Since $\xi_{-,0} >\xi_{+,0}$ beyond $r\sim 15\himpc$, $\xi_{\times\times}(r)$ is negative at such scales, as measured for haloes from simulations and galaxies from observation \citep[fig. 6 of ][]{Okumura:2009}. The II correlation function is known to be harder to measure and noisier than the GI correlation function. Moreover, the amplitude of $\xi_{\times\times}$ is even more suppressed compared to $\xi_{++}$ because of the large anisotropy \citep{Croft:2000,Okumura:2009}. Interestingly, however, the quadrupole moment of $\xi_{\times\times}$ is larger than other II correlation components. Probing the multipole moments may enable one to easily measure the II correlation function rather than focusing on the monopole alone.

\subsection{VI correlation}
Finally, we derive the simple expression of the VI correlation function. Again, by writing $k_z/k = y_{1,0}(\hat{\vk})$ and utilizing the relation between $y_{\ell m}$ and the Wigner's 3-j symbols, the resulting VI correlation function is expressed as 
\begin{align}
\xi_{v + }(\vecr) 
&=\wt{C}_1
\cos{(2\phi)}\mu (1-\mu^2)  
\Xi_{\delta\Theta,3}^{(1)}(r). 
\label{eq:vi_la_1d}
\end{align}
Another component, $\xi_{v\times}$, is also derived in the same manner as equation ({\ref{eq:vi_la_1d}), but $\cos{(2\phi})$ term is replaced with $\sin{(2\phi})$. Just like the II correlation, the VI correlation is not affected by RSD at linear order, and we omit the superscript $S$ or $R$. We plot this function as a function of $\vecr=(r_\perp,r_\parallel)$ in the right panel of Fig.~\ref{fig:contour2D_gi_ia_th_2gpc}. Although with the velocity field we can probe the structure growth at larger scales than with the density field, the BAO features in the VI correlation are much less prominent than those in the GI and II correlations. 

From equation (\ref{eq:vi_la_1d}), we can easily find non-zero multipoles which are, $\ell=1$ and $\ell=3$, and
\begin{align}
\xi_{v +,1 }(r) 
&=-\xi_{v +,3 }(r) =
\frac{2}{5}\,\wt{C}_1\,
\Xi_{\delta\Theta,3}^{(1)}(r).
\label{eq:vi_la_1d_l}
\end{align}
Thus, there is a relation similar to the case of the GI function, but here the octopole-to-dipole ratio becomes $-1$. This is shown in the upper-right panel of Fig.~\ref{fig:gi_ii_vi_multi_th_2gpc_z014} \citep[equivalent to the blue dotted curve in fig.~12 of][]{Okumura:2019}. 

\subsection{$E$ mode auto- and cross-correlations}
By analogy with weak lensing surveys, the above alignment statistics can be decomposed into gradient type ($E$ mode) and curl type ($B$ mode) components \citep{Crittenden:2002,Schneider:2006,Troxel:2015}. Since weak lensing is known to produce only $E$ mode to the lowest order, it is useful to express our formulas derived above with the ellipticities decomposed into $E$/$B$ modes. 

As shown by \cite{Blazek:2011}, in the LA model the $E$ and $B$ mode auto correlations are simply $\xi_{EE}(\vecr) =\xi_{+}(\vecr)$ and $\xi_{BB}(\vecr)=0$. The cross-correlation between galaxies and $E$ modes in real space is derived as
\be
\xi_{gE}^R(\vecr) = -\frac{2}{3}\, \wt{C}_1\,
b_{\rm g}\,  \left[ {\cal P}_0(\mu)\,\Xi_{\delta\delta,0}^{(0)}(r) 
+{\cal P}_2(\mu)\,\Xi_{\delta\delta,2}^{(0)}(r) \right]. 
\ee
Thus we have $\xi_{gE,0}^R(r)=\xi_{gE,2}^R(r)$.
The one in redshift space contains additional terms, given as
\begin{align}
\xi_{gE}^S(\vecr) &= \xi_{gE}^R(\vecr) 
+\frac{2}{105}\, \wt{C}_1\,
f \left[
-7\,{\cal P}_0(\mu)\, \Xi_{\delta\Theta,0}^{(0)}(r)
\right. \nn \\ 
&\ \ \ \left.
+5\,{\cal P}_2(\mu)\, \Xi_{\delta\Theta,2}^{(0)}(r)
+12\,{\cal P}_4(\mu)\, \Xi_{\delta\Theta,4}^{(0)}(r)
\right].
\end{align}
Finally, the cross-correlation between velocities and $E$ modes is 
\be
\xi_{vE}(\vecr) = -\frac{2}{5}\, \wt{C}_1\,
\left[ {\cal P}_1(\mu)\,\Xi_{\delta\delta,3}^{(1)}(r) 
+{\cal P}_3(\mu)\,\Xi_{\delta\delta,3}^{(1)}(r) \right]. 
\ee
For $\xi_{vE}$, the dipole and octopole moments coincides with each other, $\xi_{vE,1}(r)=\xi_{vE,3}(r)$. Beyond the leading order, the $B$-mode contribution is no longer zero \citep[See][for the modeling of the II correlation including the $B$ mode]{Blazek:2019}. In such a case, the IA statistics measured in configuration space are a mixture of $E$ and $B$ modes, just like the weak-lensing shear statistics. 

\section{Summary}\label{sec:conclusion}
We have presented new formulas for various intrinsic alignment statistics and derived their explicit angular dependences: the GI correlation in real space (equation \ref{eq:gi_la_1d}) and in redshift space (equation \ref{eq:gi_la_s_1d}), II correlation (equations \ref{eq:ii_la_1d_p} and \ref{eq:ii_la_1d_m}) and VI correlation (equation \ref{eq:vi_la_1d}). They are essential to fully extract cosmological information from BAOs and RSDs encoded in IAs of galaxies. 

Orientations of galaxies are known to be not perfectly aligned with those of the host haloes. However, the overall shape of the IA correlation is found to remain unchanged, with the amplitude to some extent reduced \citep{Okumura:2009,Okumura:2009a}. Thus, the uncertainties due to the galaxy-halo misalignments can be absorbed into the free parameter $C_1$ of the LA model (equation \ref{eq:gamma_la}) and the formulas presented in this \paper\, can be directly applied to the observed alignment statistics on large scales. 

In a companion paper \citep{Okumura:2020} we make a detailed comparison of our formulas of the IA statistics to the $N$-body simulation measurements. We show that the anisotropies measured in the simulations are accurately predicted by our anisotropic LA model, and the accuracy is further improved by extending the LA model to include the nonlinear clustering \citep[i.e., NLA model;][]{Bridle:2007}. Following the success of the LA model, in another companion paper \citep{Taruya:2020} we perform a Fisher matrix analysis of the IA statistics and demonstrate that cosmological constraints will be significantly improved compared to the case using galaxy clustering alone. Small-scale behaviors of the IAs beyond linear theory have been actively studied using nonlinear perturbation theory, and higher-order terms need to be taken into account particularly at $r<10\himpc$ \citep{Blazek:2015,Blazek:2019,Vlah:2020}. The nonlinearity not only puts an impact on the anisotropies in the IA statistics but also induces non-zero $B$-mode. Hence, as is the case with the clustering statistics, it is crucial to develop models which accurately describe the IA statistics at such scales in order to fully extract the cosmological information. On large scales, the galaxy clustering statistics are known to be affected by the wide-angle effect \cite[e.g.,][]{Szalay:1998}. Similarly, it affects the alignment statistics on such scales. These effects on the newly derived IA statistics will be investigated in our future work. 

\section*{Acknowledgments}
We thank Takahiro Nishimichi for the collaboration on related projects \citep{Okumura:2019,Okumura:2020}. T.~O. thanks Aniket Agrawal for useful discussion. 
T.~O. acknowledges support from the Ministry of Science and Technology of Taiwan under Grants No. MOST 106-2119-M-001-031-MY3 and the Career Development Award, Academia Sinina (AS-CDA-108-M02) for the period of 2019 to 2023. 
A.~T. was supported in part by MEXT/JSPS KAKENHI Grants No. JP15H05889 and No. JP16H03977.

\bibliographystyle{mnras} 

\label{lastpage}

\end{document}